\documentclass[letterpaper,fleqn,usenatbib]{mn2e}

\usepackage{times}
\long\def\symbolfootnote[#1]#2{\begingroup%
\def\thefootnote{\fnsymbol{footnote}}\footnote[#1]{#2}\endgroup}
\usepackage[T1]{fontenc}
\usepackage{ae,aecompl}

\usepackage{graphicx}	
\usepackage{amsmath}	
\usepackage{amssymb}	
\usepackage{multirow}
\usepackage{natbib}
\usepackage{subfigure}
\usepackage{xcolor}
\usepackage{ulem}
\usepackage{tabu}
\usepackage{url}

\title[Discovery of the Jet stream]{Discovery of a thin stellar stream in the SLAMS survey}

\author[P. Jethwa et al.]{P. Jethwa$^{1,2}$\thanks{E-mail: pjethwa@eso.org},
G. Torrealba$^{2,3}$,
C. Navarrete$^{4,5,2}$,
J. A. Carballo-Bello$^{4,5}$,
T. de Boer$^{2,6}$,	\newauthor
D. Erkal$^{2,6}$,
S.E. Koposov$^{7,2}$,
S. Duffau$^{5,4,8}$,
D. Geisler$^{9}$,
M. Catelan$^{4,5}$
\& V. Belokurov$^{2,10}$
\\
$^{1}$European Southern Observatory, Karl-Schwarzschild-Str. 2, 85748 Garching, Germany\\
$^{2}$Institute of Astronomy, University of Cambridge, Madingley Road, Cambridge CB3 0HA, UK\\
$^{3}$Institute of Astronomy and Astrophysics, Academia Sinica, P.O. Box 23-141, Taipei 10617, Taiwan\\
$^{4}$Instituto de Astrof\'isica, Pontificia Universidad Cat\'olica de Chile, Av. Vicu{\~{n}}a Mackenna 4860, 782-0436 Macul, Santiago, Chile\\
$^{5}$Millennium Institute of Astrophysics, Av. Vicu{\~{n}}a Mackenna 4860, 782-0436 Macul, Santiago, Chile\\
$^{6}$Department of Physics, University of Surrey, Guildford, GU2 7XH, UK \\
$^{7}$McWilliams Center for Cosmology, Carnegie Mellon University, 5000 Forbes Avenue, Pittsburgh, PA 15213, USA\\
$^{8}$Departamento de Ciencias Fisicas, Universidad Andres Bello, Fernandez Concha 700, Las Condes, Santiago, Chile\\
$^{9}$Departamento de Astronomia, Universidad de Concepcion, Casilla 160-C, Concepcion, Chile\\
$^{10}$Center for Computational Astrophysics, Flatiron Institute, 162 5th Avenue, 10010, New York, NY, USA
}

\date{Accepted XXX. Received YYY; in original form ZZZ}

\pubyear{2017}

\begin{document}
\label{firstpage}
\pagerange{\pageref{firstpage}--\pageref{lastpage}}
\maketitle

\begin{abstract}
We report the discovery of a thin stellar stream - which we name the Jet stream - crossing the constellations of Hydra and Pyxis.
The discovery was made in data from the SLAMS survey, which comprises deep $g$ and $r$ imaging for a $650$ square degree region above the Galactic disc performed by the CTIO Blanco + DECam.
SLAMS photometric catalogues have been made publicly available.
The stream is approximately 0.18 degrees wide and 10 degrees long, though it is truncated by the survey footprint.
Its colour-magnitude diagram is consistent with an old, metal-poor stellar population at a heliocentric distance of approximately 29 kpc.
We corroborate this measurement by identifying a spatially coincident overdensity of likely blue horizontal branch stars at the same distance.
There is no obvious candidate for a surviving stream progenitor.
\end{abstract}




\section{Introduction}
\label{sec:intro}

Stellar streams are the relics of satellite galaxies or globular clusters disrupting in the tidal field of their host \citep{newberg16}, predicted by hierarchical theories of structure formation whereby galaxies grow through mergers of smaller units \citep{lyndebell95}.
Roughly a dozen examples have been discovered in the Milky Way (MW), supporting hierarchical growth models and, in addition, offering a way to answer two crucial questions about the MW:
What is the large-scale mass distribution of its dark matter halo \citep{johnston99}?
Does it contain low-mass dark matter subhalos \citep{ibata_etal_2002,johnston_etal_2002}?
Though streams have provided the strongest constraints to-date on the halo mass distribution \citep[e.g.][]{koposov_etal_2010,law_majewski_2010,gibbons_sgr_2014,bowden15,kuepper_etal_2015,bovy16}, the existence of low-mass subhalos remains an open, important question.

The $\Lambda$CDM cosmological model predicts that the MW halo contains hundreds of thousands of subhalos \citep{diemand08,springel08}, the majority of which are predicted to be too low-mass to host any baryonic counterpart \citep{ikeuchi86,rees86}.
Thin stellar streams in the Galactic halo may provide the first evidence for such a population, the basic idea being that a subhalo interaction can induce an observable density perturbation -- a gap -- in the stream.
Though the theory behind gap formation is well developed \citep{yoon_etal_2011,carlberg12,erkal15a,sanders_etal_2016}, a difficulty in unambiguously establishing the existence of low-mass subhalos is that streams in the inner-halo may also be perturbed by giant molecular clouds \citep{amorisco16} or by the Milky Way bar \citep{erkal17,pearson_etal_2017}. Both of these possibilities are a relevant concern for the gaps and density perturbations discovered in the Pal 5 stream \citep{carlberg_pal5_2012,bovy_pal5_2017,erkal17}. Finding thin stellar streams in the outer halo, where interloping perturbers are less common, is therefore an important step to test an outstanding prediction of $\Lambda$CDM.

We recently carried out a mini survey which led to the fortuitous discovery of a thin stellar stream in the outer halo, which we name the Jet stream.
The survey, dubbed the Search for the Leading Arm of Magellanic Satellites (SLAMS), was proposed to look for satellites of the Magellanic Clouds, testing a prediction from \citet{jethwa16} that some Magellanic satellites should reside in a leading arm, preceding the Clouds.
No such satellites were discovered, however, a null result which will be discussed in a future work.
In this work we present the SLAMS survey and the stream discovery: we describe the survey and data reduction in Section 2, present the stream and its properties in Section 3, discuss possible stream progenitors in Section 4, then summarise our conclusions in Section 5.

\section{Data}
\label{sec:data_slams}

\begin{figure}
\centering
\includegraphics[width=\columnwidth]{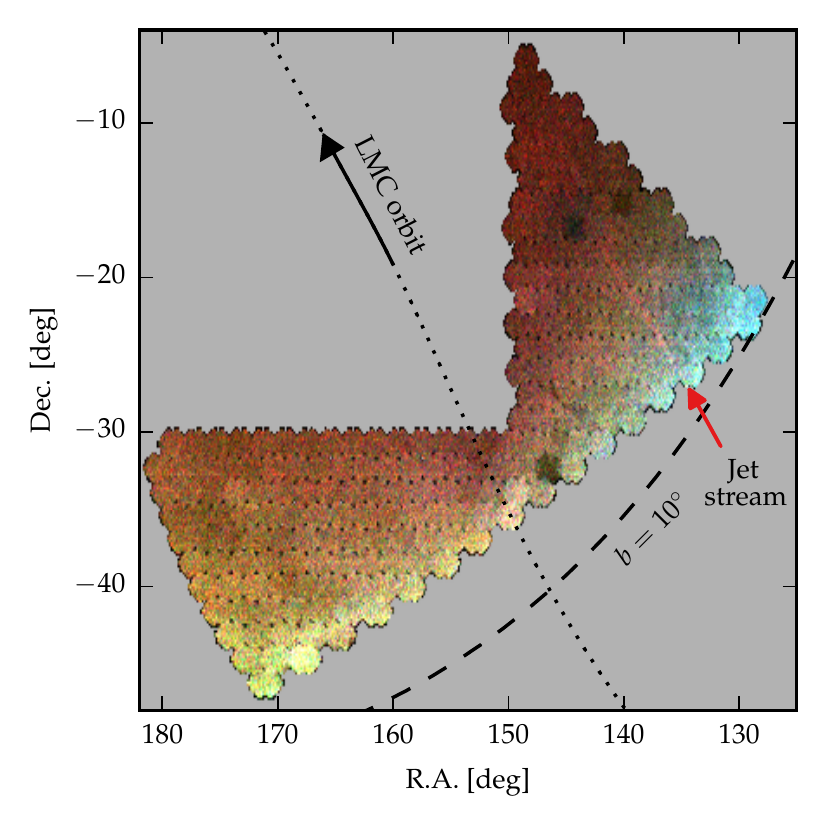}
\caption[False-colour stellar density image]{
False-colour map of the density of stars with $0.1<g-r<0.45$ from the SLAMS survey.
Density in different magnitude ranges is shown using different colour channels: red $20.5<r<23.0$, green $17.5<r<20.5$ and blue $15.0<r<17.5$.
The newly discovered stream is labelled.
The dotted line shows the future orbit of the LMC, calculated using \citet{kallivayalil13} proper motions, which is roughly aligned with the orientation of the Jet stream.
The green/blue colour gradient along the southern edge is due to the varying distance distribution of MW disk stars.
}
\label{fig:false_color}
\end{figure}

SLAMS is a $650$ deg$^2$ optical survey conducted using DECam on the 4-m Blanco telescope at Cerro Tololo Inter-American Observatory in Chile.
DECam's 2.2 degree field of view is amongst the widest available for ground-based optical imaging, making it well suited to efficiently image the desired region.
The survey footprint was designed to enclose the predicted location of Magellanic Cloud satellites \citep{jethwa16}, and it is bordered on the southern edge by a line of roughly constant Galactic latitude $b=10^\circ$, and on the northern edge by the footprint of the VST ATLAS survey \citep{shanks15}.
Roughly half the region is covered by the Pan-STARRS survey \citep{kaiser02} however, as we shall show in Section~\ref{ssec:spop}, this did not reach sufficient depth for our requirements.

Data were taken over four half nights in the 2016B semester, between 30-31 December 2016 and 30-31 January 2017.
The survey footprint was covered with a single, un-dithered tiling, consisting of one 90 second exposure in the DES $g$-band and a co-located 90 second exposure in the DES $r$-band.
These exposure times were chosen to reach a $10\sigma$ limiting magnitude of $\sim22.5$ mag in both filters, only $0.5$ mag shallower than the depth achieved in the first two years of DES data.
The survey consists of $\sim500$ exposures over an area of $650$ deg$^2$, with $13\%$ of the area covered by multiple exposures, giving an almost complete coverage ($>95\%$) of the footprint.

Observing conditions were variable, with seeing in the range 0.8-2.2'', with a median of 1.0''.
Ten exposures suffered seeing conditions worse than 1.6'' and enhanced sky brightness due to close-to-twilight observation times.
Other than these, the desired depth was achieved for the rest of the survey.
Processed images and weight maps were downloaded from the NOAO Science Archive.
These processed images, delivered by the DECam pipeline, have been bias, dark and flat-field calibrated as well as cross-talk corrected, whilst WCS astrometry is also provided.

To extract catalogues from the processed images we followed the recipe described in \citet{koposov15} which was used to process DES images, and makes extensive use of the \texttt{SExtractor} and \texttt{PSFEx} routines \citep{bertin96,bertin11,annunziatella13}.
The steps taken were to (i) run \texttt{SExtractor} on the image files with the provided weight maps for fast source extraction, (ii) generate a model of the point spread function (PSF) by running \texttt{PSFEx} on each CCD chip, (iii) run \texttt{SExtractor} again, now with the PSF model, to determine \texttt{PSF\_MAG} magnitudes, (iv) ingest the resulting catalogues into a PostgreSQL database and create \texttt{Q3C} spatial indices \citep{koposov06} to speed up spatial querying, and (v) remove duplicate catalogue entries from overlapping fields by performing a $1''$ auto-cross-match and retaining only the entry with minimum photometric error.

Photometric calibration was performed by cross-matching between catalogues from SLAMS and APASS DR9 \citep{henden15}, restricting to $m_\mathrm{APASS}>16.5$ mag to avoid saturated stars.
Zero points were calculated per-field as the median photometric offset with respect to APASS, then an additional correction was made per CCD chip by combining all fields and calculating the median offset per chip.
We estimate the photometric precision by comparing duplicate measurements of an object taken from overlapping exposures.
The resulting photometric precision is $45$ mmag, as defined by taking the best-fitting Gaussian, $\sigma$, to the distribution of photometric offsets between duplicate pairs.

The final source catalogue\footnote{available at \url{zenodo.org/record/1344449}} is compiled by combining $g$ and $r$ bands using a $1''$ matching radius.
We derived extinction corrected magnitudes using DES bandpass-specific extinction coefficients from \citet{schlafly11}, assuming a $R_V=3.1$ reddening law, and the associated $E(B-V)$ reddening values from \citet{schlegel98} interpolated at the position of each star.
All magnitudes shown henceforth are extinction corrected.
We perform star-galaxy separation using the criterion that stars must satisfy
\begin{equation}
|\texttt{SPREAD\_MODEL}| < 0.003 + |\texttt{SPREADERR\_MODEL}|,
\end{equation}
in both $g$ and $r$ bands .
This criterion is shown in \citet{koposov15} to provide a good balance between completeness and purity at the faint end.

Figure~\ref{fig:false_color} illustrates the quality of the resulting catalogue.
It shows a density map of stars in the colour range $0.1<g-r<0.45$, where the red, green and blue image colour channels correspond to magnitude ranges $20.5<r<23.0$, $17.5<r<20.5$ and $15.0<r<17.5$ respectively.
Some defects are apparent.
A handful of fields suffered from poor observing conditions, most clearly visible as the $\sim2^\circ$ diameter dark patches centred at (R.A.,Dec.) = $(146^\circ,-32^\circ)$ and $(145^\circ,-18^\circ)$.
Additionally, sub-degree sized inter-field gaps are visible throughout the entire footprint, a consequence of our decision to perform un-dithered exposures in order to maximise sky coverage.
Aside from these defects the density distribution appears uniform, attesting to the survey's good photometric stability and robust star-galaxy separation.
The blue and green regions along the southern edge of the image show the increasing contribution of nearby disk stars.

\section{The Jet Stream}
\label{sec:jet1}

The Jet stream was fortuitously discovered through visual inspection of SLAMS stellar density maps.
It is visible in Figure~\ref{fig:false_color} as the linear overdensity, whose red colour indicates that the stream stars preferentially lie at faint magnitudes.
For the initial characterisation, we therefore adopt a correspondingly faint colour-magnitude selection box to enhance the stream signal: $0.1<g-r<0.4$, $21<r<23$.
We will also transform into a great circle co-ordinate system $(\phi_1,\phi_2)$ aligned with the stream.
To find this, we use a least squares optimisation routine to determine the pole which maximises the number of SLAMS stars with latitude $|\phi_2|<1^\circ$ which lie inside the colour magnitude box given above.
This pole is given in Table~\ref{tab:jet1_properties}.
With no obvious progenitor to set the zero-point, we choose $\phi_1=0$ to intersect the line R.A.=$138.789^\circ$, crossing the centre of the detected section of stream.
With no obvious stream progenitor to set the lonigitude zero-point, we place $\phi_1=0$ at the center of the detected portion of stream, namely (R.A., Dec.)=$(138.789^\circ, -21.903^\circ)$.

\begin{figure}
\centering
\includegraphics[width=\columnwidth]{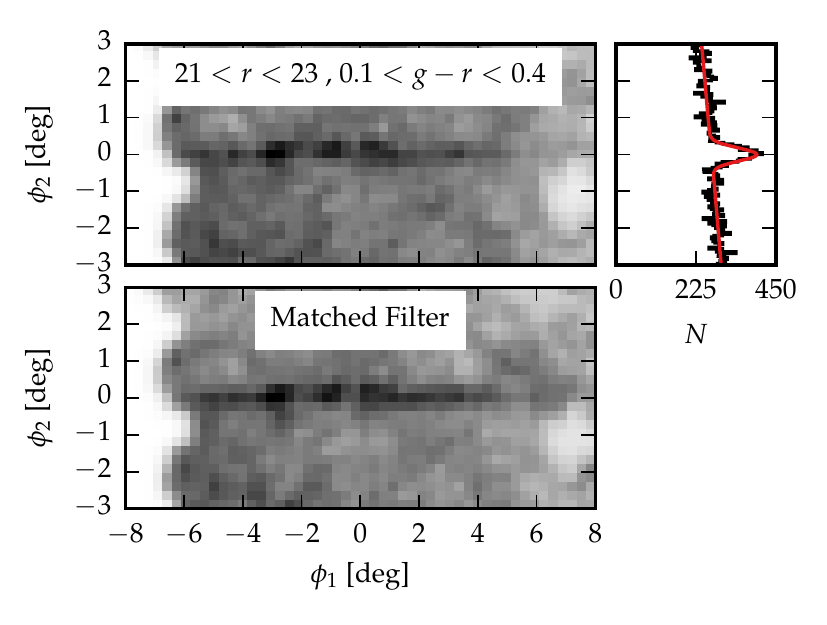}
\caption[Density in stream co-ordinates]{
Jet stream density.
Top left panel shows SLAMS stellar density in the given colour magnitude box, rotated into stream co-ordinates.
The black histogram in the top right panel shows star counts integrated along $\phi_1$ between $[-6^\circ,5.5^\circ]$, the red line shows a maximum likelihood model consisting of a linear background model and a Gaussian with a width of $\sigma_{\phi_2}=0.18^\circ$.
The bottom panel shows the density after applying a matched filter based on the best fit isochrone.
}
\label{fig:rotate_stream}
\end{figure}

Figure~\ref{fig:rotate_stream} shows the transformed stellar density map.
The top left panel shows density using the simple colour magnitude selection.
The right panel shows star counts integrated over $\phi_1$, with the stream appearing as a significant overdensity at $\phi_2=0^\circ$.
We model the integrated stellar density as the sum of a linear background and a Gaussian component centred at $\phi_2=0^\circ$,
\begin{equation}
n(\phi_2) = p_0 + p_1 \phi_2 + N \frac{1}{\sqrt{2\pi\sigma_{\phi_2}^2}} \exp\left( -\frac{1}{2}\left(\frac{\phi_2-0}{\sigma_{\phi_2}}\right)^2 \right).
\label{eqn:phi2mod}
\end{equation}
We fit this model to stars with $\phi_1$ in $[-6^\circ,5.5^\circ]$, avoiding the range $[5.5^\circ,6^\circ]$ where there is significant incompleteness.
We infer a width $\sigma_{\phi_2}=(0.18\pm0.02)^\circ$ and $N=1620\pm200$ stars comprising the stream.
The central values are the maximum likelihood solution, found using downhill simplex optimisation, and the uncertainties are calculated via Markov Chain Monte Carlo sampling of the posterior probability density function (PDF) assuming uniform uninformative priors.
These basic properties are summarised in the top section of Table~\ref{tab:jet1_properties}.

\begin{table}
	\centering
	\footnotesize
	\caption[Stream properties]{
	Stream properties.
	Central values and error-bars (where given) correspond to maximum likelihood values and the $95\%$ credibility intervals of the posterior PDF.
	The top section shows basic properties, the middle section shows results from CMD fits, and the bottom section shows other derived constraints.
	}
	\label{tab:jet1_properties}
	\begin{tabu}{lll}
	\hline
	Property							&	Value 															& Unit 				\\
	\hline
	$\phi_1$ range					&	$-6^\circ$ to $5.5^\circ$												& -					\\
	Stream start (R.A., Dec.)	&	$(134.671^\circ, -26.584^\circ)$							& -					\\
	Stream end (R.A., Dec.)		&	$(142.329^\circ, -17.526^\circ)$							& -					\\
	Pole	(R.A., Dec.)				&	$(64.983^\circ,34.747^\circ)$									& -					\\
	Width (Gaussian $\sigma$, on-sky)	&	$0.18\pm0.02$													& deg 				\\
	\hline
	$m-M$								&	$17.28^{+0.05}_{-0.07}$											& mag 				\\
	$M_*$ 								&	$25\pm2$														& $10^3 M_\odot$	\\
	Age 								&	$12.1^{+0.9}_{-0.3}$					& Gyr				\\
	$\log Z/Z_\odot$		&	$-1.7^{+0.1}_{-0.3}$					& -					\\
	\hline
	Distance 							&	$28.6^{+0.7}_{-0.9}$											& kpc				\\
	Width (Gaussian $\sigma$, physical)	&	$90\pm10$														& pc				\\
	\hline
	\end{tabu}
\end{table}

\subsection{Stellar population}
\label{ssec:spop}

\begin{figure}
\centering
\includegraphics[width=\columnwidth]{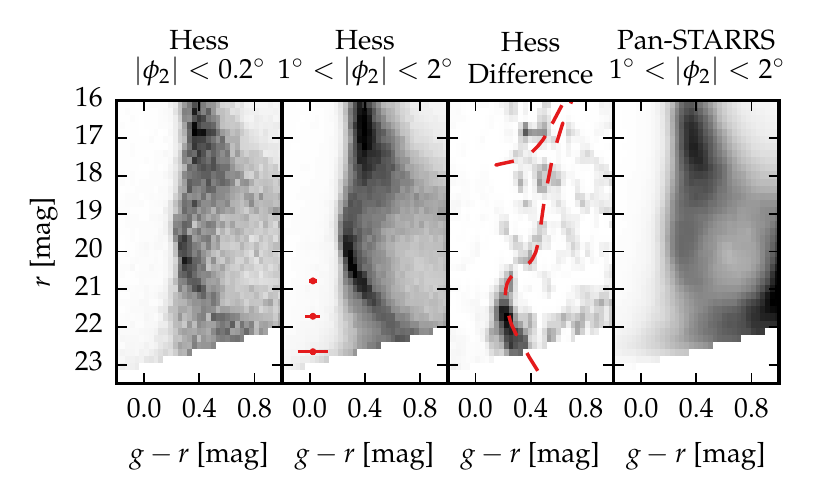}
\caption[Stream Hess diagrams]{
Stream Hess diagrams.
The first three panels show SLAMS data, for (from left to right) a region on-stream, a background region, and the Hess difference between these.
The Hess difference shows a clear MSTO well described by an old, metal-poor isochrone at 29 kpc (shown in red, details given in the text).
Median photometric errors are shown in red in the second panel.
The right panel shows a Hess diagram from Pan-STARRS for the same region as the second panel, highlighting the superior data quality of SLAMS.
}
\label{fig:hess}
\end{figure}

We next characterise the stellar population of the stream.
Figure~\ref{fig:hess} shows Hess diagrams for a region on the stream (first panel, $|\phi_2|<0.2^\circ$) and away from it (second panel, $1^\circ<|\phi_2|<2^\circ$, i.e. both above and below the stream), each normalised by sky area.
The Hess difference between these regions (third panel) shows an unmistakable overdensity corresponding to a main sequence turn-off (MSTO) and, slightly less prominently, a sub-giant branch.
These can be well described by a theoretical isochrone (dashed red line) with metallicity $\log Z/Z_\odot=-1.6$, age $12.1$ Gyr and distance modulus $m-M=17.3$ mag.
We will describe the steps taken to derive these parameters below, however here we note that there are some features visible in the Hess difference that are not well described by the best-fitting isochrone.
The sub-giant branch and tentative red giant branch lie blue-wards of the model, possibly indicating a problem with synthetic DECam magnitudes used to construct the isochrone \citep[as previously noted in][]{simon15}.
The origin of a horizontal stripe visible in the Hess difference at $r=17$ mag is also unclear.

To highlight the quality of SLAMS data, in the rightmost panel of Figure~\ref{fig:hess}, we also show a Hess diagram from the Pan-STARRS survey corresponding to the same region of sky as the second panel.
Though unrelated to the stream, note the well defined main sequence (MS) and MSTO evident between $19<r<22$ in SLAMS data.
This corresponds to the Monoceros ring, an outer-disk structure in the MW.
The Monoceros ring is discernible in Pan-STARRS data \citep{morganson16} but, as can be seen in the rightmost panel, in this region of sky the sharpness of this structure is washed out by Pan-STARRS' photometric errors.
Similarly, the overdensity in the bottom-right corner of the Pan-STARRS' Hess diagram consists of nearby disk dwarfs scattered bluewards by photometric errors.
The enhanced depth of SLAMS fixes this issue, and furthermore explains why the Jet stream had not been discovered in the previous survey.

\begin{figure}
\centering
\includegraphics[width=\columnwidth]{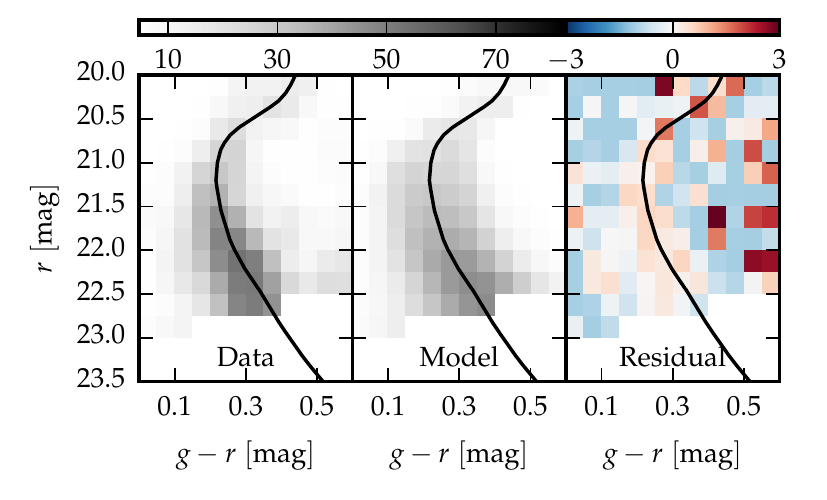}
\caption[Fitting the stream MSTO]{
Fitting the stream MSTO.
The left panel shows the stream-weighted CMD from SLAMS data, the central panel shows the maximum likelihood synthetic CMD.
The value of the colour-map for these panels, shown above, corresponds to the absolute number of stars inside each pixel.
The right panel shows the model-normalised residual, i.e. (data - model) / model.
The maximum likelihood isochrone is reproduced in each panel.
}
\label{fig:turnoff}
\end{figure}

Although the Hess difference in Figure~\ref{fig:hess} already shows a strong stream signal in colour magnitude space, this signal can be further enhanced by weighting stars according to their probability of belonging to the stream, rather than separating stream stars using a hard cut in $\phi_2$.
Enhancing the signal this way will be especially important when we later come to analyse the stream in $\phi_1$ bins.
Here, we first construct a stream-weighted colour magnitude diagram (CMD) for the entire length of the stream.
In each CMD pixel, we fit the model given by Equation~(\ref{eqn:phi2mod}) to the $\phi_2$ distribution of stars in that pixel, but now fixing the width to the value derived earlier, i.e. $\sigma_{\phi_2}=0.18^\circ$.
We use the downhill simplex optimisation to find the maximum likelihood values of the remaining parameters, with the stream-weighted CMD given by the number of stars in the Gaussian component of the model in each pixel.
The stream-weighted CMD of the MSTO is shown in the left panel of Figure~\ref{fig:turnoff}.

We now derive the metallicity, age, and distance of the Jet stream.
The outline for how we do this is to (i) take a grid of isochrones, (ii) construct synthetic CMDs, which we then (iii) fit against the observed, stream-weighted CMD.
For the first step, we take a grid of theoretical isochrones \citep{bressan12} spanning a range of ages (6-13 Gyr), metallicities ($\log Z/Z_\odot$ in $[-2.1,-1]$), and assuming a Chabrier initial mass function \citep[IMF;][]{chabrier11}.
These are provided in filters appropriate for DECam.
We also grid over distance moduli ($m-M$ in $[15,20]$) and total stellar mass of the stream ($\log M_*/M_\odot$ in $[3,5]$).
For each of these parameters we take a grid of 50 values spaced linearly in the quoted ranges.

Secondly, we transform isochrones into synthetic CMDs by convolving them with the photometric uncertainty of our data.
In each CMD pixel we calculate the median photometric errors of all SLAMS stars in that pixel in the vicinity of the stream, which we call $\sigma_g$ and $\sigma_r$.
Assuming these are uncorrelated, the error covariance matrix in $(g-r,r)$ space is given by
\begin{equation}
\mathrm{Cov}_{(g-r,r)} = \begin{pmatrix} \sigma_g^2+\sigma_r^2 & -\sigma_r^2 \\ -\sigma_r^2 & \sigma_r^2 \end{pmatrix}.
\end{equation}
Stepping along an isochrone we distribute the predicted number stars throughout the CMD according to the error covariance matrix.
We do this using isochrone step sizes smaller than the chosen CMD pixel scale, to avoid unwanted discretisation in the synthetic CMD.
Given the lack of available and comparably deep photometry in the SLAMS region, we were unable to easily perform completeness tests.
Rather than model incompleteness, for initial characterisation we therefore simply restrict our CMD fits to $g<23$, which is only 0.5 mag shallower than our presumed $10\sigma$ limiting magnitude.

Finally, we construct a likelihood function in order to probabilistically compare the synthetic CMDs to the observed, stream-weighted CMD.
Given the observed, stream-weighted CMD $d(g-r,r)$ and the synthetic, model CMD $m(g-r,r)$, we define a Poisson likelihood function,
\begin{equation}
\mathcal{L} = \prod_{i,j} \frac{m_{ij}^{d_{ij}}\exp(-m_{ij})}{d_{ij}!}
\end{equation}
where the indices $i,j$ run over the CMD pixels.
We calculate the logarithm of this likelihood function over the grid of model parameters described above.

Figure~\ref{fig:turnoff} shows the maximum likelihood solution when we fit the stream-weighted CMD of the MSTO.
The model (middle panel) is a reasonable reproduction of the data (left panel), though there is clear room for improvement since the shape of the MSTO can be discerned in the residuals (right panel).
The mismatch may be explained if our stream-weighting procedure systematically over/under-estimates the stream signal in CMD pixels near-to/far-from the best-fitting isochrone.
Given that the MSTO appears tighter in the data than in the model, however, we speculate that this mismatch is more likely to be due to an overestimation of photometric errors.
The maximum likelihood isochrone has metallicity $\log Z/Z_\odot=-1.6$, age $12.1$ Gyr and distance modulus $m-M=17.3$ mag.
We calculate the uncertainties from the posterior distribution with uniform priors on the parameters.
The middle section of Table~\ref{tab:jet1_properties} lists the maximum likelihood values and $95\%$ confidence intervals for parameters describing the stream's stellar population.
Our constraints on the stream age and metallicity exhibit a well known degeneracy \citep{worthey94}, both of which also display a degeneracy with stream distance.
Covariances between all other parameters are small.

We have restricted this fit to the MSTO since other regions of the CMD, especially at brighter magnitudes, contain significant contamination from non-stream stars.
This can be seen in the Hess Difference shown in Figure~\ref{fig:hess}.
When we do attempt to fit the entire CMD, we typically find that a small number of stars along the asymptotic giant and red giant branches of the isochrone contribute $\Delta \log \mathcal{L} \approx +10$ to the likelihood, essentially breaking the age-metallicity degeneracy.
Though these stars may be genuine stream members, given the level of background contamination at $r<20$ and our lack of realistic background model, such a conclusion is very uncertain.
In order to avoid spuriously shrinking our derived parameter uncertainties, we therefore chose to fit only the MSTO.

Our constraints on distance modulus correspond to heliocentric distance constraints of $28.6^{+0.7}_{-0.9}$ kpc, i.e. the maximum likelihood
value and 95\% statistical confidence intervals.
These uncertainties do not include any systematics in isochrone fitting.
At this distance, the apparent width of the stream corresponds to a physical width of $90\pm10$ pc.
This morphologically places the Jet stream in the category of thin stellar streams along with e.g. Pal 5 \citep{odenkirchen03}, GD-1 \citep{grillmair06}, and ATLAS \citep{koposov14}.
The thinness is suggestive of a globular cluster progenitor for the stream; we discuss this issue further in Section~\ref{sec:prog}.

\subsection{Matched filter selection}

Having found an isochrone that well describes the stream's stellar population, we can use it to select stream stars more accurately than was achieved using the simple colour-magnitude selection box.
To do this, we weight stars by the ratio of their probability of stream membership to their probability of belonging to a background population, as a function of their colour and magnitude i.e.
\begin{equation}
\frac{P_\mathrm{stream}(g-r,r)}{P_\mathrm{BG}(g-r,r)}.
\end{equation}
The synthetic CMD generated from the maximum likelihood isochrone found in the previous section defines $P_\mathrm{stream}(g-r,r)$.
The background PDF $P_\mathrm{BG}(g-r,r)$ is found empirically from on-sky regions adjacent to the stream, following e.g. \citet{rockosi02,koposov10}.

The bottom panel of Figure~\ref{fig:rotate_stream} shows the resulting image after CMD-weighting the stream.
Compared to the simple colour-magnitude selection, the stream signal is clearly enhanced, and apparent over a greater range of $\phi_1$.
The extent of the stream is still curtailed, however, on one end by the SLAMS footprint, and on the other by a region of poor data quality and hence it may well extend past these limits.

\subsection{Distance along the stream}
\label{ssec:distvar}

We next characterise the variation of distance along the stream length.
This is done by fitting synthetic isochrones as described in Section~\ref{ssec:spop} for eight evenly spaced bins for $\phi_1$ in $[-6.5^\circ,5.5^\circ]$.
In contrast to Section~\ref{ssec:spop}, we now fix age and metallicity of the isochrone to the maximum likelihood values determined previously, as we do not expect these quantities to vary over the stream length.
Although there is a significant degeneracy between these parameters, we have checked that the qualitative behaviour of the distance variation we infer from the maximum likelihood age and metallicity is robust against sampling different values along this degeneracy.

\begin{figure}
\centering
\includegraphics[width=\columnwidth]{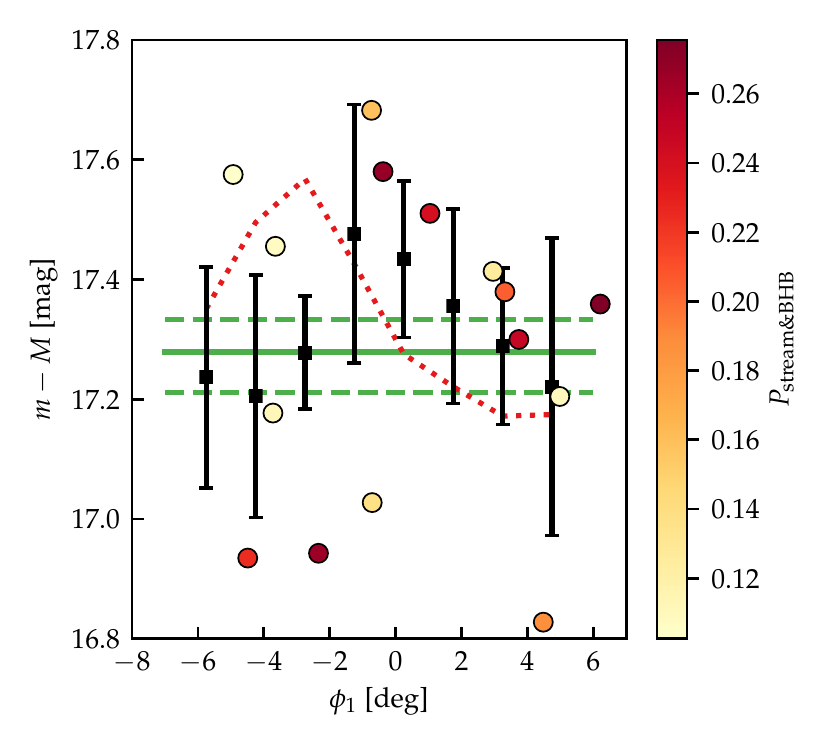}
\caption[Variation of distance along the stream]{
Variation of distance modulus along the stream.
Green horizontal lines show the median value and 95\% credibility intervals inferred from isochrone fits to the stream's MSTO.
Black squares with error bars show the results from fits in eight bins of $\phi_1$.
Coloured circles represent stream BHB candidates coloured by a metric related to the probability of being a genuine BHB (based on a Pan-STARRS colour-colour cut), and stream association (based on $\phi_2$).
Typical BHB distance modulus uncertainties are $0.15$ mag (not shown).
The dotted red line indicates the variation of extinction along the stream: it shows the median $A_V$ in the region $|\phi_2|<0.3^\circ$ for eight $\phi_1$ bins, offset by 17 mag.
}
\label{fig:distance}
\end{figure}

Figure~\ref{fig:distance} shows the results.
The square data points with error-bars show the median and 95\% credible intervals of the stream distance modulus in $\phi_1$ bins from fitting the MSTO CMD.
The results are consistent at $2\sigma$ significance with a constant distance throughout the observed region, however the central two measurements lie slightly further away than the outer six.
This may indicate a turning point in the stream distance, possibly corresponding to the orbital apocenter of the stream progenitor.

To scrutinise the purported turning point, in Figure~\ref{fig:distance} we also show extinction along the stream as the red dotted line.
$A_V$ varies by $\sim0.4$ mag along the stream length, peaking at $\phi_2\sim-3^\circ$.
Whilst the extinction variation shows some qualitative similarities to our inferred distance variation, it is different in detail: the $\phi_1$ of peak extinction and peak distance differ by $1.5^\circ$, and the drop in extinction towards negative $\phi_1$ is less severe than the drop in stream distance.
Therefore, whilst systematics due to extinction may contribute at some level to the measured distance variation, it is difficult to see how it could explain the full behaviour.
To further investigate the issue, we next turn to an alternative distance estimator.

\subsection{Blue horizontal branch selection}

All of the constraints we have presented thus far on the stream's distance rely on CMD fitting to the stream's MSTO.
Such CMD fits are subject to a number of systematic uncertainties arising from e.g. uncertainties in the IMF, background modelling, CMD binning, binary stars, extinction, as well as intrinsic systematic errors in the model isochrones we have used.
Whilst it is difficult to control all of these systematics, we can complement and corroborate the results from CMD fits using an alternative stellar tracer.

Blue horizontal branch (BHB) stars are an eminently popular tracer for studies of the Galactic halo and structures therewithin \citep{yanny00,newberg03,xue08,bell10,deason11,ruhland11}, thanks to a number of useful properties.
Their peculiar colours make them easily identifiable, their absolute magnitudes \citep[$M_g\sim0.5-0.7$, e.g.][]{sirko04} allow detection to large distances, and their small spread in absolute magnitude \citep[$\sigma_{M_r}\sim 0.15$][]{deason11} make them one of the best distance estimators available, outperformed only by RR Lyrae stars.

\begin{figure}
\centering
\includegraphics[width=\columnwidth]{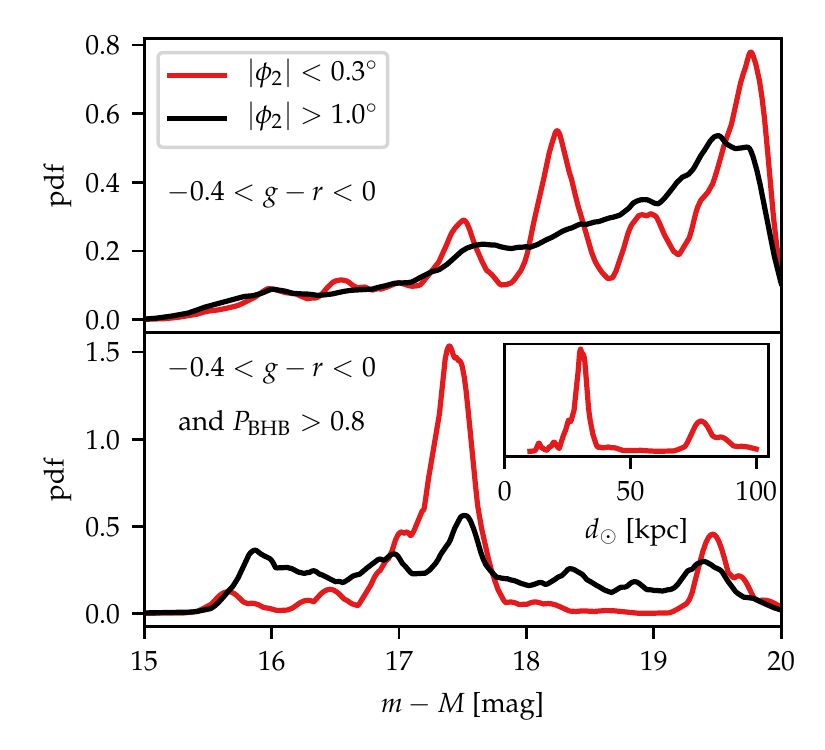}
\caption[Detection of the Jet stream in BHBs]{
Detection of the Jet stream in BHBs.
In each panel, lines show a KDE of the PDF of distance moduli assuming a sample consisting of BHBs with absolute magnitudes given by the formula in \citet{belokurov16}.
Red lines show the distribution for a region on-stream, black for the background.
The top panel shows this for all stars with $g-r$ in $[-0.4,0]$, with the stream showing multiple low-significance overdensities compared to the background.
In the bottom panel we restrict to the most likely BHB candidates using a selection based on Pan-STARRS colours shown in Figure~\ref{fig:bhb_select}.
The purer BHB sample brings out a significant overdensity at $m-M\approx17.4$ mag, corresponding to the Jet stream.
}
\label{fig:bhbs}
\end{figure}

As a first attempt to identify stream BHBs we make a simple CMD selection of stars with $g-r$ in $[-0.4,0]$ and $r<21$.
We use the \citet{belokurov16} BHB absolute magnitude calibration \citep[which is a slightly modified version of that given in][]{deason11} to estimate the absolute magnitude of these stars as function of their $g-r$ colour.
This colour is a proxy for stellar temperature, the primary determinant of BHB luminosity.
The top panel of Figure~\ref{fig:bhbs} shows the resulting distribution of inferred BHB distance moduli, for a region on the stream (red line, $|\phi_2|<0.3^\circ$, 261 stars) and a background region away from it (black line, $1^\circ<|\phi_2|<4^\circ$, 2132 stars).
The PDFs are kernel density estimates (KDE) using an Epanechnikov kernel with a bandwidth $0.25$ mag, normalised to have unit area in the range of distance modulii shown.
The chosen bandwidth is equal to the range of the eight median constraints of the stream's distance modulus from the MSTO, which we adopt as a prior estimate of the intrinsic spread in distance modulus along the stream.
Examining Figure~\ref{fig:bhbs}, we see that the stream distribution shows three peaks at $m-M>17$, all of roughly comparable height above the background.
There is little sign of any strong stream signal.

We hypothesised that contamination may be the cause of this failure to detect the stream in BHBs since quasars, white dwarfs, blue straggler and young main-sequence stars can all overlap the range of $g-r$ colour occupied by BHBs.
To remedy this, we need a purer BHB sample.
Were spectroscopy available, small differences in the shapes of the Balmer lines would provide a powerful way to identify BHBs from other stars with similar temperature \citep{sirko04}.
Accumulated over the whole Balmer series, these differences can even be detected in broad-band photometry \citep{yanny00}, but this requires high-quality $u-$band photometry, which is not available in SLAMS or any other coincident survey.
Thankfully, near-IR photometry can also identify BHBs \citep{vickers12}, with only marginally less success than the near-UV, where the discriminating power now arising from differences in the line-shape of the Paschen, rather than Balmer, series.
Though unavailable in SLAMS, we make use of the near-IR $i-$ and $z-$band photometry from Pan-STARRS to purify our BHB selection.

\begin{figure}
\centering
\includegraphics[width=\columnwidth]{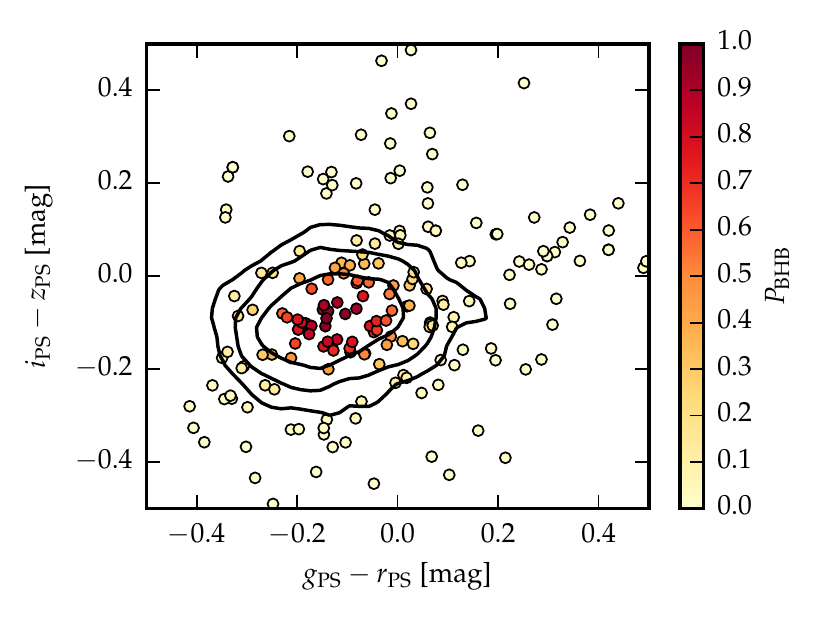}
\caption[BHBs selection using Pan-STARRS]{
Selection of BHBs using Pan-STARRS photometry.
Black contours represent the density of spectroscopically confirmed BHB stars from \citet{xue11} in Pan-STARRS colour-colour space.
Circles show a random sample of SLAMS stars with $-0.4<g-r<0$ and $r<21$.
These are coloured by their proximity to the spectroscopically confirmed sample, $P_\mathrm{BHB}$.
}
\label{fig:bhb_select}
\end{figure}

Figure~\ref{fig:bhb_select} shows distributions in Pan-STARRS $(g-r,i-z)$ colour-colour space.
The black contours show the distribution of spectroscopically confirmed BHBs from \citet{xue11}, cross matched to Pan-STARRS using a $1''$ matching radius.
We model their probability density in colour-colour space as a bivariate normal with the mean and covariance of the data.
Next we cross match SLAMS with Pan-STARRS, and assign each SLAMS star a value $P_\mathrm{BHB}$ based on the model of the \citet{xue11} BHBs, assigning $P_\mathrm{BHB}=1$ to a star lying at the mean the \citet{xue11} sample.
Note that the \citet{xue11} BHBs have very similar magnitude range to our objects of interest, hence we can safely compare their colour distributions.
The circles in Figure~\ref{fig:bhb_select} show a random selection of 200 SLAMS stars with $g-r$ in $[-0.4,0]$ and $r>21$, coloured by $P_\mathrm{BHB}$.
There is a concentration of SLAMS stars within the innermost black contour which are likely to be genuine BHBs, while other stellar types display a broader distribution throughout colour-colour space.

Note that the sample shown in Figure~\ref{fig:bhb_select} contains stars with $g-r<0$ in SLAMS but $g-r\sim0.4$ in Pan-STARRS.
Visually inspecting images of these objects suggests that many have line-of-sight close neighbours, and hence are possibly binary stars with variable photometry.
By selecting BHB candidates based on SLAMS $g-r$ colour then making an additional selection using Pan-STARRS $g-r$ colour, we remove these contaminants, but note that it is the $i-z$ colour which primarily distinguishes BHBs.
We have checked that our subsequent analysis does not significantly change if we restrict our BHB selection only to SLAMS $g-r$ and Pan-STARRS $i-z$ colours.

The bottom panel of Figure~\ref{fig:bhbs} shows the distribution of BHB distance moduli after restricting our selection to stars with $P_\mathrm{BHB}>0.8$.
The resulting distribution of on-stream stars (red line, 33 stars) exhibits a single large peak above the background level (black line, 203 stars).
All other peaks visible in the top panel of Figure~\ref{fig:bhbs} have been greatly diminished in significance after applying the $P_\mathrm{BHB}$ cut, and hence were likely the result of contamination.
We can unambiguously identify the one remaining peak with the Jet stream.
This peak is centered at $17.4$ mag - which is reassuringly consistent with the mean distance of the stream inferred from the MSTO - and spans distance modulii $[16.8,17.8]$, a range which contains 15/33 of the on-stream sample.
We will use this range of distance modulii as a criterion to select likely stream BHBs, but note that this range itself is not a reliable estimate of the line-of-sight depth of the stream, since it is inflated by (i) the KDE smoothing kernel, and (ii) a significant number of background contaminants (discussed below).

\begin{figure}
\centering
\includegraphics[width=\columnwidth]{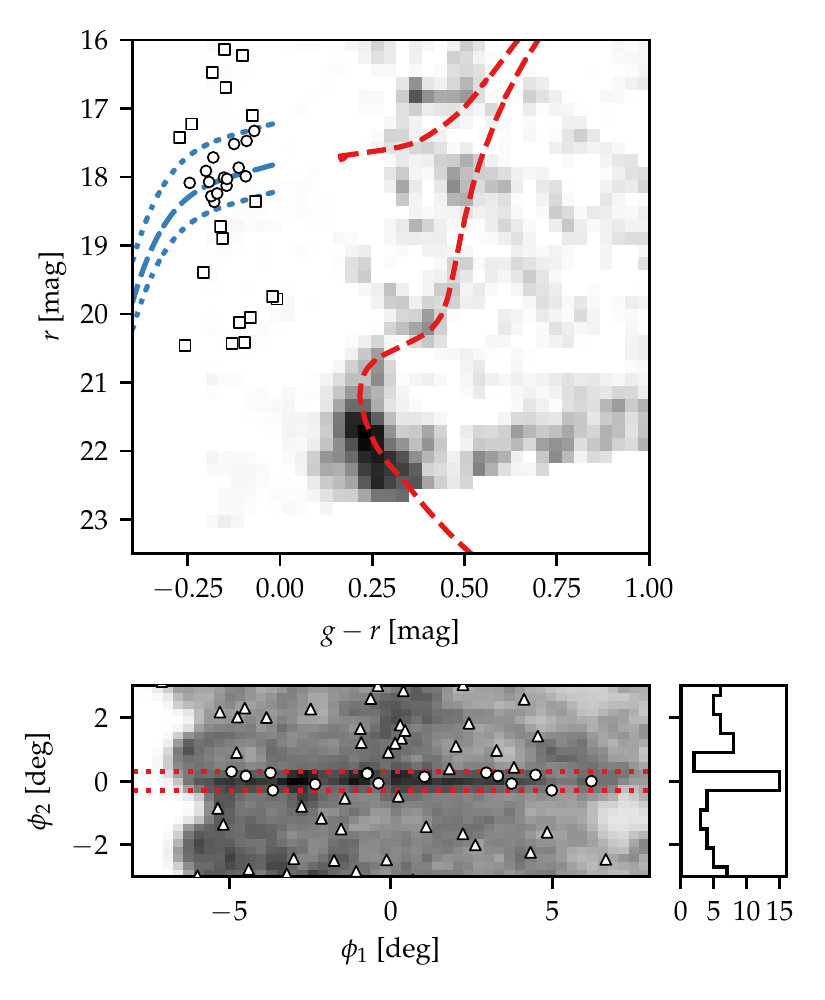}
\caption[Distribution of BHBs]{
Distribution of $P_\mathrm{BHB}>0.8$ BHB candidates.
The top panel shows distribution in colour-magnitude space: white symbols show candidates with $|\phi_2|<0.3^\circ$, the dashed blue line shows our BHB ridge-line at $m-M=17.4$ mag, dotted lines show ridgelines at $16.8/17.8$ mag, whilst the grey-scale distribution and red dashed line are the stream Hess difference and isochrone repeated from Figure~\ref{fig:hess}.
In the bottom panel, overlaid on the matched filter image from Figure~\ref{fig:rotate_stream}, white symbols show the on-sky distribution of BHB candidates with $16.8<m-M<17.8$, whilst dotted red lines delineate the region $|\phi_2|<0.3^\circ$.
The histogram on the right shows the distribution collapsed along $\phi_1$, with a significant peak at $|\phi_2|<0.3^\circ$.
}
\label{fig:bhb_cmd_spatial}
\end{figure}

In Figure~\ref{fig:bhb_cmd_spatial}, we show the distribution of stream BHB candidates, selected using the cut $P_\mathrm{BHB}>0.8$.
The top panel shows colour-magnitude distribution of candidates with $|\phi_2|<0.3^\circ$ (white symbols) overlaid on the stream Hess difference (grey-scale) and isochrone fit to the MSTO (dashed red line).
We can see many likely BHBs aligning with the BHB ridge-line $m-M=17.4$ mag (dashed blue line).
These give rise to the single large peak in the red PDF in the bottom panel of Figure~\ref{fig:bhbs}.
In the bottom panel, white symbols show the on-sky distribution of BHB candidates with distance moduli in $[16.8,17.8]$.
Collapsed along $\phi_1$ (histogram on right), their distribution shows a single significant peak within $|\phi_2|<0.3^\circ$.

We add a sample of likely BHBs associated with the stream to Figure~\ref{fig:distance}.
This sample consists of all stars with $g-r<0$, $r>21$, $P_\mathrm{BHB}>0.8$, $m-M$ in $[16.8,17.8]$, and $|\phi_2|<0.3^\circ$ (i.e. the white circles in Figure~\ref{fig:bhb_cmd_spatial}).
Note that these particular cuts in $P_\mathrm{BHB}$ and $|\phi_2|$ were chosen as they maximise the stream BHB signal in Figure~\ref{fig:bhbs}, however a strong signal remains when we vary their values by $\pm50\%$.
We colour the stars added to Figure~\ref{fig:distance} by
\begin{equation}
P_\mathrm{stream\&BHB} = P_\mathrm{BHB} P_\mathrm{stream}(\phi_2),
\end{equation}
where $P_\mathrm{BHB}$ is as described earlier, and
\begin{equation}
P_\mathrm{stream}(\phi_2) = \frac{N \frac{1}{\sqrt{2\pi\sigma_{\phi_2}^2}} \exp\left( -\frac{1}{2}\left(\frac{\phi_2}{\sigma_{\phi_2}}\right)^2 \right)}{n(\phi_2)}
\end{equation}
is the fraction of stars at a given $\phi_2$ associated with the stream from Equation~(\ref{eqn:phi2mod}), where we take the maximum likelihood values of parameters found earlier.

Examining Figure~\ref{fig:distance}, we see a sequence of likely stream BHBs with $\phi_1$ in $[-1^\circ,6^\circ]$ exhibiting distances which decrease with increasing $\phi_1$, in line with the trend inferred from the CMD fit for this portion of the stream.
For the region $\phi_1<-1^\circ$, there are fewer likely stream BHBs.
Two candidates at $\phi_1<-1^\circ$ and $m-M\sim16.9$ have relatively large values of $P_\mathrm{stream\&BHB}$, and hence must have Pan-STARRS colours consistent with being a BHB, and lie close to the stream equator.
Given that they are inconsistent with the MSTO inference, however, these may be background contaminants.
Six BHB candidates in Figure~\ref{fig:distance} are inconsistent at $2\sigma$ with the MSTO inference, including two at $\phi_1<-1^\circ$ and $m-M\sim16.9$ with relatively large values of $P_\mathrm{stream\&BHB}$.
To check whether this number of discrepancies can be explained by background contaminants, we estimate the number of MW halo BHBs contributing to Figure~\ref{fig:distance} from neighbouring on-sky regions.
The sample shown in Figure~\ref{fig:distance} numbers 15: this is equal by construction to the central bin of the histogram in Figure~\ref{fig:bhb_cmd_spatial}.
From the adjacent histogram bins we estimate the number of MW halo BHB contaminants at $5.0\pm1.7$.
It is therefore plausible that the 6 BHB candidates discrepant with the MSTO in Figure~\ref{fig:distance} are background contaminants.
Excluding these possible contaminants, the spread in distance modulus inferred from the MSTO and BHBs is broadly consistent, at $\sim0.4$ mag, corresponding to $\sim5.5$ kpc at the stream's distance.

We also note that the number of likely BHBs we have identified in the stream is consistent with its measured stellar mass.
The MSTO fit gives a stellar mass for the stream of $2.5\times10^4M_\odot$ which, assuming a stellar mass-to-light ratio appropriate for an old, metal-poor population, corresponds to an absolute magnitude $M_V\sim-5$.
MW satellites with $M_V\sim-5$ have between 10-20 BHB stars - e.g. the globular cluster Palomar 15 \citep{harris91} and three dwarf galaxies \citep{sand12} - in line with the numbers presented for the Jet stream.

Considered altogether, the mean distance of the stream ($\sim29$ kpc) inferred from the MSTO and that inferred from BHBs strongly support one another.
For $\phi_1$ in $[-1^\circ,6^\circ]$, the distance variation of the stream as measured from the MSTO and BHBs agree well, with both tracers indicating a $\sim5.5$ kpc change in this $\phi_1$ range.
Given that the on-sky stream width is only $\sim90$ pc it is very likely that this distance change is due to the orientation of the stream - i.e. due to the orbit of the progenitor - rather than intrinsic stream thickness.
For $\phi_1<-1^\circ$, BHBs are poorly sampled and hence no firm conclusion can be drawn on whether the stream has a distance turning point near $\phi_1\sim0^\circ$.

\section{The progenitor?}
\label{sec:prog}

What was the progenitor of the Jet stream?
An extrapolation of its great circle passes just $1.1^\circ$ from the Sextans dwarf spheroidal, however given that this dwarf lies approximately 50 kpc further away than the visible portion of the stream, this alignment may very well be coincidental.
Other satellites within $4^\circ$ of the great circle extrapolation include the dwarfs Carina, Reticulum 2 and Horologium 1, and the globular clusters Palomar 3 and NGC 1261. Proper motions measurements, however, would be required to make a more conclusive statement regarding the progenitor.

Despite not being able to pin down the identity of the progenitor, can we say more certainly whether it was likely a cluster or a dwarf galaxy?
We infer a stream width of $90\pm10$ pc and stellar mass is $2.5\pm0.2\times10^4M_\odot$.
Assuming a stellar mass-to-light ratio appropriate for an old, metal-poor population, this corresponds to an absolute magnitude $M_V\sim-5$.
Equating these stream properties to the properties of its progenitor, then placed on the size-luminosity relation \citep[see Figure 10 of][for a recent version]{torrealba16b}, the Jet stream progenitor would overlap the distribution of dwarf galaxies, not globular clusters.
This simple argument would suggest that the stream progenitor may have been a dwarf galaxy.

A problem with this argument is that there is a non-trivial mapping between the progenitor size and the stream width.
This is shown by the analytic model of stream evolution from \citet{esb_16}, which successfully reproduces the widths of streams in $N$-body simulations.
In this model, stream width is not governed by the progenitor size alone, but instead by the ratio of velocity dispersion at the tidal radius of the progenitor to its orbital velocity at pericenter.
\citet{erkal16} simplify the \citet{esb_16} model by assuming that the host has a logarithmic potential and the progenitor is on a circular orbit.
This results in a simple expression for the dynamical mass of a stream progenitor as a function of stream width and the mass of the host enclosed inside the stream's galactocentric radius \citep[equation 28 of][]{erkal16}.
Applying this expression to the Jet stream, which has $r_\mathrm{GC}\sim32$ kpc, and taking $M(<32\;\mathrm{kpc})\sim2.8\times10^{11}M_\odot$ \citep[e.g.][]{williams15}, the model predicts that the dynamical mass of the stream progenitor was $6.4\times10^4M_\odot$.
Dividing this by the estimate of the stellar mass of the stream gives a mass to light ratio of $2.6$.
This is in-line with expectations for globular clusters, but an order of magnitude below the expected value for a dwarf galaxy of the appropriate luminosity.
According to this calculation, it is therefore likely that the progenitor of the Jet stream was a globular cluster.

We next consider whether the Jet stream may be associated with other known MW streams by comparing against the recent compilation of \citet{mateu18}.
Based on orientations of on-sky stream tracks, the only possible association could be with PS1-B \citep{bernard16}, which is less than $10^\circ$ away on-sky and has an orbital pole $2.24^\circ$ away.
A simple association is unlikely, however, since PS1-B is at a heliocentric distance of 14.5 kpc compared to Jet at 28.6 kpc - directly connecting the two would require a distance gradient much more extreme than we have measured.
Alternatively, Jet and PS1-B could be (i) successive wraps of the same stream around the MW if the progenitor has undergone substantial orbital decay, (ii) formed from two distinct globular clusters which were accreted as part of a larger group, or (iii) simply co-incidentally aligned.
Radial velocities and chemical abundances of stream stars may help to discriminate these possibilities.

Lastly, we consider whether the stream progenitor may have been associated with the LMC.
The dotted line in Figure~\ref{fig:false_color} shows the predicted future orbit of the LMC, calculated in a MW potential comprising a $10^{12}M_\odot$ NFW halo and using the \citet{kallivayalil13} kinematics.
Comparison of this orbit with the track of the Jet stream shows that the two are roughly aligned.
As a first approximation, a stream traces the orbit of its progenitor \citep[though this is not true in detail,][]{sanders13}.
With the Jet stream and the LMC orbit roughly aligned, it is therefore plausible that the progenitor was an LMC satellite (so long as the LMC has a sufficiently large halo).
Though by no means a proof of association, this shows that this scenario is plausible.
The fact that the stream lies along a great circle at all, however, may be suggestive that the stream formed around the MW, not the LMC.
Whichever scenario is true, it will be the case that the stream is one of the thin stellar streams closest to the LMC.
Indeed, if it extends southwards, beyond the edge of the SLAMS survey footprint, it may be the closest such stream.
Given that the LMC can perturb stellar streams \citep{veraciro13}, this raises the exciting prospect that the Jet stream could be used to constrain the highly uncertain total mass of the LMC  \citep{penarrubia16}.

\section{Conclusions}
\label{sec:concs}

We have presented the SLAMS survey, and reported the discovery of a thin stellar stream in SLAMS data.
The stream properties derived in this work are:
\begin{itemize}
\item it lies on a great circle with pole at (R.A., Dec.)$=(64.983^\circ,34.747^\circ)$,
\item it has heliocentric distance $\sim 29$ kpc, constrained using both MSTO and BHB stars,
\item its width on the sky is 0.18 degrees corresponding to $\sim90$ pc physical size,
\item its CMD resembles that of an old, metal-poor isochrone.
\end{itemize}
A full description of the main properties we derive is given in Table~\ref{tab:jet1_properties}.

We envisage a succession of future observations to further characterise the stream.
Additional imaging is planned to attempt to trace the stream beyond the current survey footprint, followed by a spectroscopic campaign to determine radial velocities, metallicities, and detailed abundances, shedding light on the nature and orbital history of the progenitor.
Finally, deeper, uniform imaging along the stream track will be required to robustly detect density perturbations caused by possible subhalo encounters.

\section*{acknowledgements}

PJ thanks the Science and Technology Facilities Council for the award of a studentship.
We gratefully acknowledge support from CONICYT/RCUK's PCI programme through CONICYT grant DPI20140066.
CN acknowledges  CONICYT-PCHA grant Doctorado Nacional 2015-21151643.
SD acknowledges support from Comit\'{e} Mixto ESO-Gobierno de Chile.
The research leading to these results has received funding from the European Research Council under the European Union's Seventh Framework Programme (FP/2007-2013) / ERC Grant Agreement n. 308024.
Additional support for this project is provided by the Chilean Ministry for the Economy, Development, and Tourism's Millennium Science Initiative through grant IC\,120009, awarded to the Millennium Institute of Astrophysics (MAS); by the Basal Center for Astrophysics and Associated Technologies (CATA) through grant PFB-06/2007; by FONDECYT grant \#1171273; and by CONICYT-Chile FONDECYT Postdoctoral Fellowship \#3160502.

This project used data obtained with the Dark Energy Camera (DECam), which was constructed by the Dark Energy Survey (DES) collaboration. Funding for the DES Projects has been provided by the U.S. Department of Energy, the U.S. National Science Foundation, the Ministry of Science and Education of Spain, the Science and Technology Facilities Council of the United Kingdom, the Higher Education Funding Council for England, the National Center for Supercomputing Applications at the University of Illinois at Urbana-Champaign,
the Kavli Institute of Cosmological Physics at the University of Chicago, the Center for Cosmology and Astro-Particle Physics at the Ohio State University, the Mitchell Institute for Fundamental Physics and Astronomy at Texas A\&M University, Financiadora de Estudos e Projetos, Funda{\c c}{\~a}o Carlos Chagas Filho de Amparo {\`a} Pesquisa do Estado do Rio de Janeiro, Conselho Nacional de Desenvolvimento Cient{\'i}fico e Tecnol{\'o}gico and the Minist{\'e}rio da Ci{\^e}ncia, Tecnologia e Inovac{\~a}o, the Deutsche Forschungsgemeinschaft,
and the Collaborating Institutions in the Dark Energy Survey. The Collaborating Institutions are
Argonne National Laboratory, the University of California at Santa Cruz, the University of Cambridge, Centro de Investigaciones En{\'e}rgeticas, Medioambientales y Tecnol{\'o}gicas-Madrid, the University of Chicago, University College London, the DES-Brazil Consortium,
the University of Edinburgh, the Eidgen{\"o}ssische Technische Hoch\-schule (ETH) Z{\"u}rich,
Fermi National Accelerator Laboratory, the University of Illinois at Urbana-Champaign,
the Institut de Ci{\`e}ncies de l'Espai (IEEC/CSIC), the Institut de F{\'i}sica d'Altes Energies, Lawrence Berkeley National Laboratory, the Ludwig-Maximilians Universit{\"a}t M{\"u}nchen and the associated Excellence Cluster Universe, the University of Michigan,
{the} National Optical Astronomy Observatory, the University of Nottingham,
the Ohio State University, the University of Pennsylvania, the University of Portsmouth,
SLAC National Accelerator Laboratory, Stanford University, the University of Sussex,
and Texas A\&M University.

Based on observations at Cerro Tololo Inter-American Observatory, National Optical Astronomy Observatory (CNTAC Prop. CN2016B-115; PI C. Navarrete), which is operated by the Association of Universities for Research in Astronomy (AURA) under a cooperative agreement with the National Science Foundation.

We thank the referee for a detailed and constructive report.





\footnotesize{ \bibliographystyle{mn2e} \bibliography{mybib} }





\bsp	
\label{lastpage}
\end{document}